\documentclass[reprint,
superscriptaddress,
nofootinbib,
 amsmath,amssymb,
 aps,
a4paper,twocolumn, floatfix
]{revtex4-2}
\usepackage{dsfont}
\usepackage{booktabs}
\usepackage{pgfplots}
\usepackage{subcaption}
\usepackage[labelfont=bf,
   justification=Justified,
   format=plain]{caption}
\usepackage{physics}
\usepackage{graphicx}
\usepackage{dcolumn}
\usepackage{comment}
\usepackage{bm}
\usepackage[hidelinks]{hyperref}
\usepackage{xcolor}
\usepackage[version=4,arrows=pgf-filled,
textfontname=sffamily,
mathfontname=mathsf]{mhchem}
\hypersetup{
    colorlinks,
    linkcolor={red!50!black},
    citecolor={blue!50!black},
    urlcolor={blue!80!black}
}

\newcommand{\shiftleft}[2]{\makebox[0pt][r]{\makebox[#1][l]{#2}}}

\date{\today}

\begin{document}

\title{Spectral Gap Superposition States}
\author{Davide Cugini}
 \email{davide.cugini01@universitadipavia.it}
 \author{Francesco Ghisoni}
 \email{francesco.ghisoni01@universitadipavia.it}
 \author{Angela Rosy Morgillo}
 \email{angelarosy.morgillo01@universitadipavia.it}
\author{Francesco Scala}
 \email{francesco.scala01@universitadipavia.it}
\affiliation{%
Dipartimento di Fisica, Universit\`a degli Studi di Pavia, via A. Bassi 6, 27100 Pavia (Italy)
}%

\begin{abstract}
    This work introduces a novel NISQ-friendly procedure
    for estimating spectral gaps in quantum systems. 
    By leveraging Adiabatic Thermalization,
    we are able to create the Spectral Gap Superposition state, 
    a newly defined quantum state
    exhibiting observable fluctuations in time 
    that allow for the accurate 
    estimation of any energy gap. 
    Our method is tested 
    by estimating the energy gap 
    between the ground and the first excited state 
    for the 1D and 2D Ising model,
    the Hydrogen molecule \ce{H2} and Helium molecule \ce{He2}.
    Despite limiting our circuit design to have at most
    40 Trotter steps, our numerical experiments of both noiseless and noisy devices for the presented systems give relative errors in the order of $10^{-2}$ and $10^{-1}$. Further experiments on the IonQ Aria device lead to spectral gap estimations with a relative error of $10^{-2}$ for
    a 4-site Ising chain, demonstrating the validity of the procedure for NISQ devices and charting a path towards a new way of calculating energy gaps.
\end{abstract}

\maketitle

\begin{figure*}
    \centering
    \includegraphics[trim={4.cm 12cm 4.cm 9cm},clip,width=.75\textwidth]{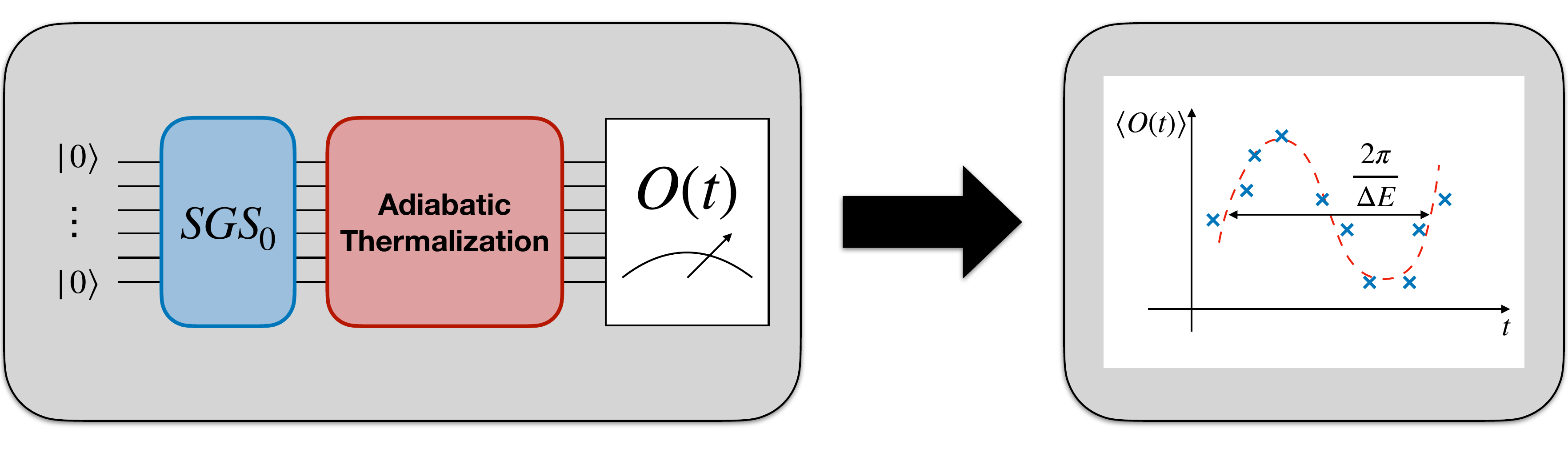}
    \caption{Schematic representation of the proposed methodology. Initially, a quantum circuit generates a Spectral Gap Superposition (SGS$_0$) state of two eigenstates of the auxiliary Hamiltonian $H_0$. Through Adiabatic Thermalization this state is evolved to the SGS state of the system Hamiltonian $H$. Sequential measurements of the chosen observable $O(t)$ at different time steps are recorded. Finally a fitting process is used to determine the spectral gap. The obtained value is compared to the benchmark spectral gap.
 }
    \label{fig:scheme}
\end{figure*}


\section{Introduction}
The estimation of the spectral gap, 
defined as the energy difference between the ground state 
and the first excited state of a quantum system, 
is fundamental in both condensed matter physics
and quantum chemistry.
Its numerical calculation on classical computers has two major hurdles: the sign problem for fermionic systems \cite{troyer2005computational}
and the exponentially increasing computational cost 
with the system size.
The advent of Quantum Computation has
given researchers tools to potentially 
overcome these fundamental issues.
Currently, we are in the so-called 
Noisy Intermediate-Scale Quantum (NISQ) era~\cite{preskill2018nisq}, 
in which devices have limited number of noisy qubits.
In fact, researchers have sought ways to 
employ these devices to come up with 
estimation techniques for the energy gap.
Some examples include: 
imaginary time propagation~\cite{leamer2023spectral},
and robust phase estimation~\cite{russo2021robustqpe}.

In this work, 
we introduce a novel approach 
for estimating the energy gaps in quantum systems. 
We define the Spectral Gap Superposition (SGS) state as
\begin{equation}
\label{eq:SGS}
\ket{SGS(i,j)} = \frac{1}{\sqrt{2}}\left(\ket{\Omega_i}  + \ket{\Omega_j} \right),
\end{equation}
where $\ket{\Omega_i}$ and $\ket{\Omega_j}$
respectively are the i-th and j-th eigenstates
of the system Hamiltonian $H$.
Although the two eigenstates are not known analytically,
they can be prepared on a quantum computer
through Adiabatic Thermalization (AT)~\cite{born1928beweis}
if the adiabatic theorem conditions hold \cite{jansen2007bounds}.
In this regime we show that it is possible to utilize
the time fluctuations of observables 
evaluated on the SGS state
to obtain an estimate of the energy gap $\Delta E_{j,i}$
between the energies of $\ket{\Omega_j}$ and $\ket{\Omega_i}$.\\

This report is organized as follows. 
In Section~\ref{sec:SGestimation} we present our procedure, 
showing its derivation and underlying principles.
In Section~\ref{sec:results} 
we report the numerical results 
of our method applied for the spectral gap estimation
of the Ising model,
on a classical computer simulator,
followed by the outcomes 
obtained with the real hardware IonQ Aria.
It follows a detailed description of the spectral gap estimation
for the \ce{H2} and \ce{He2} molecules
run on a classical computer simulator.
Finally, in Section~\ref{sec:discussion}, 
we provide a comprehensive summary 
and draw conclusions from our findings. 

\section{Spectral gap estimation}
\label{sec:SGestimation}
Let $O(t)$ be a time dependent operator of type
\begin{equation}
 O(t)   =  e^{iHt} O e^{-iHt}
\end{equation}
where $H$ is the Hamiltonian of a physical system.
The expectation value of the observable $O(t)$ calculated on the SGS state~\eqref{eq:SGS} is
\begin{equation}\label{eq: fluctuations}
\begin{split}
    \langle O(t) \rangle &= \frac{1}{2}\left( \bra{\Omega_i} O \ket{\Omega_i} + \bra{\Omega_j} O \ket{\Omega_j} \right) + \\
&\quad\quad + \rho\,\mathrm{cos}\left(\Delta E_{ji}t + \theta\right),
\end{split}
\end{equation}
where $\rho > 0$ and $\theta \in [0, 2\pi)$
are the polar coordinates
of the complex number $ \bra{\Omega_j} O \ket{\Omega_i}.$
From Equation~\eqref{eq: fluctuations} 
one could obtain an estimate of the energy gap
by fitting the $\langle O(t) \rangle$ function
over the time variable.\\
To obtain $\langle O(t) \rangle$ one should 
(i) prepare the SGS state on a quantum computer,
(ii) implement the time evolution $e^{-iHt}$ 
as a quantum circuit and 
(iii) compute the expectation value of an operator $O$ 
(see Figure~\ref{fig:scheme}).
The state preparation can be performed 
via the AT.
This allows to evolve the $n$-th eigenstate $\ket{\omega_n}$
of an auxiliary Hamiltonian $H_0 $ 
into the $n$-th eigenstate of $H$ 
through the evolution operator $U_\tau$, i.e.
\begin{equation}
    \lim_{\tau \to \infty}U_\tau\ket{\omega_n} = \ket{\Omega_n},
\end{equation}
where $\tau > 0$ represents the thermalization time.
This guarantees that if one is able to prepare the superposition 
\begin{equation}
    \ket{SGS_0} = \frac{1}{\sqrt{2}}\left(\ket{\omega_0}+ \ket{\omega_1} \right) 
\end{equation}
then  it is enough to thermalize it to get
\begin{equation}
    \lim_{\tau \to \infty}U_\tau\ket{SGS_0} = \ket{SGS}.
\end{equation}
The AT evolution, 
like the time evolution operator of step (ii),
can be implemented through the \textit{Trotter-Suzuki method} \cite{Trotter}.

It is important to note that,
as long as $\rho>0$,
any observable $O$ can be used in Equation \eqref{eq: fluctuations}
to estimate the energy gap.
\textit{A priori} knowledge of the system should be used to find the observable that maximizes $\rho$, since this facilitates the fitting process.
Observe that our procedure allows,
in principle, 
to estimate the energy gap between two arbitrary eigenstates of $H$.
Moreover, this is done without separately calculating the two energy eigenvalues,
keeping the overall computational cost 
as low as possible.

\section{Numerical results}
\label{sec:results}
In this section, we present the numerical tests 
with the developed procedure.  
In the perspective of implementing the 
numerical studies on a Quantum device,
we restrict the circuits to be composed of 
40 Trotter steps.
From now on we use the following notation
for the $2\times2$ identity operator 
and the Pauli matrices
\begin{equation}
    \sigma^0 = I, \quad \sigma^1 = \sigma^x,
    \quad \sigma^2 = \sigma^y,
    \quad \sigma^3 = \sigma^z.
\end{equation}

\subsection{Ising model}
\label{sec:Ising}
Our first test case is the $L$-sites Ising model \cite{huang2008statistical}
with Periodic Boundary Conditions (PBC), 
described by the Hamiltonian 
\begin{equation}
    H = -\frac{J_1}{2} \sum_{\langle ij \rangle} \sigma^1_i \sigma^1_{j} - \frac{h_3}{2}\sum_{i}\sigma^3_i,
    \label{eq:ising}
\end{equation}
where $\sigma_i$ is a Pauli matrix acting on site $i$
and the $\langle \cdot \cdot  \rangle$ symbol 
restricts the sum over the nearest neighbors.
In what follows, we restrict to the case 
of positive coupling constants
$h_3 ,J_1 \geq 0$.
To find the spectral gap
we need to prepare the SGS state
\begin{equation}
\label{eq:sgsising}
\ket{SGS(0,1)} = \frac{1}{\sqrt{2}} (\ket{\Omega_0}+ \ket{\Omega_1}).
\end{equation}
For this scope we consider the auxiliary Hamiltonian
\begin{equation}
    H_0 = -\frac{J_1}{2}\sum_{\langle ij \rangle } \sigma^1_i\sigma^1_{j}.
\end{equation}
The ground state of $H_0$ space is degenerate
with dimension 2 and is spanned by 
$\ket{+}^{\otimes L} $ and $\ket{-}^{\otimes L} $,
which are defined as:
\begin{equation}
    \ket{\pm} \equiv \frac{1}{\sqrt{2}}\left(\ket{0} \pm \ket{1} \right).
\end{equation}
Preparing $\ket{SGS_0}$ requires the right 
linear combination of 
$\ket{+}^{\otimes L} $ and $\ket{-}^{\otimes L} $.
From symmetry considerations,
which are reported in Appendix~\ref{Ising SGS},
we find
\begin{equation}
    \ket{SGS_0(0,1)} =  \ket{+}^{\otimes L}.
\end{equation} 
All experiments conducted for the Ising model,
both in 1D and 2D, use the observable
\begin{equation}\label{eq:observableising}
    O =  \sigma^3 \otimes \left(\sigma^0\right)^{\otimes (L-1)}.
\end{equation}
Numerical experiments
demonstrate that this observable maximizes $\rho$ 
in Equation~\eqref{eq: fluctuations} 
for any Ising chain (1D case) and square lattice (2D case). 
The full investigation can be found in 
in Appendix~\ref{sec:observable}.

The presented experiments are a 
4 site Ising chain with PBC.
It is important to note that
the circuit depth per Trotter step needed to simulate Ising 
is independent of the number of sites $L$. 
In particular, all experiments 
involving the 4-site Ising chain, 
including noiseless, noisy and real hardware, 
have a maximum depth of 80 2-qubit gates
once composed with IonQ 
native gates. 
More information can be seen in Appendix~\ref{subsec:ionq}.

The 40 Trotter steps
are divided in 15 thermalization steps 
and 25 time evolution steps. We choose the 25 times $t$ for the evaluation of $O(t)$ 
to be Chebyshev-distributed 
to facilitate the fitting process \cite{Rendon2024improvedaccuracy}
(see Appendix~\ref{sec:chebyshev}).
Each circuit
is simulated on a classical computer 
with 8192 shots.
We take the intrinsic error 
of the quantum state measurement 
into account as described in Appendix~\ref{sec: Intrinsic error}. 
All simulated results are compared
to a numerical benchmark which is 
calculated using direct diagonalization of 
the Hamiltonian $H$.

The numerical results for the 4-site Ising chain, presented in Figure 
\ref{fig:spectral gap running}, 
show an excellent agreement
between the numerical benchmark 
and the noiseless estimated spectral gap for $L=4$. 
The graph illustrates the spectral gap 
dependence on the parameter $h_3/J_1$. 
The relative error between the noiseless 
numerical results and the benchmark is in the order of $10^{-2}$.

The infrared and adiabatic limits are regimes where our approach becomes unstable,
within the imposed number of Trotter steps\footnote{One can always try to balance the reduced number of Trotter steps with longer time steps, but at a certain point, the Trotter-Suzuki approximation would no longer hold.}.
In the infrared limit, the $O(t)$ frequency tends to zero 
and its estimation would require a huge amount of time steps for the time evolution. 
The adiabatic limit, instead,  requires long thermalization time for a high fidelity state preparation. 

Remarkably, our approach exhibits strong resilience 
when subjected to the noise model inspired by IonQ Aria 1. 
Specifically, the impact of thermal relaxation and readout are minimal due to the extended $T_1$ and $T_2$ parameters and small readout error, while the biggest error contribution is due to 2-qubit gate average fidelities (see Appendix~\ref{subsec:ionq} for more details on the hardware noise model). 
The average relative error between the numerical and the estimated gap has been determined to be of order $10^{-1}$.
Moreover, further investigation revealed that the procedure has also strong
resilience to reduced number of shots (see Appendix~\ref{subsec:ionq}).

Analogous results are obtained for the 4-site Ising lattice 
and are reported in Appendix \ref{subsec:Ising2d}. 

\subsubsection*{IonQ Aria Device}
The final test
for the SGS procedure on a 4 
site Ising chain 
is the IonQ Aria 1 device.
Due to resource limitations,
the conducted experiments 
used slightly different specs compared
 to the noiseless and noisy experiments.
Firstly, the energy gap for only 3 different 
$h_3/J_1$ values is found. 
These values are selected 
since they are far from both the infrared and adiabatic limits.
Secondly, only a subset of 10 times $t$ of the initial 25, 
are employed for the fitting of $O(t)$. 
Finally, 2500 shots (with error mitigation) 
are used instead of 8192 employed in the numerical 
simulations (without error mitigation).

\begin{figure}[]
    \centering 
    \includegraphics[trim={.5cm .5cm 0.cm 0cm},clip,width=.45\textwidth]{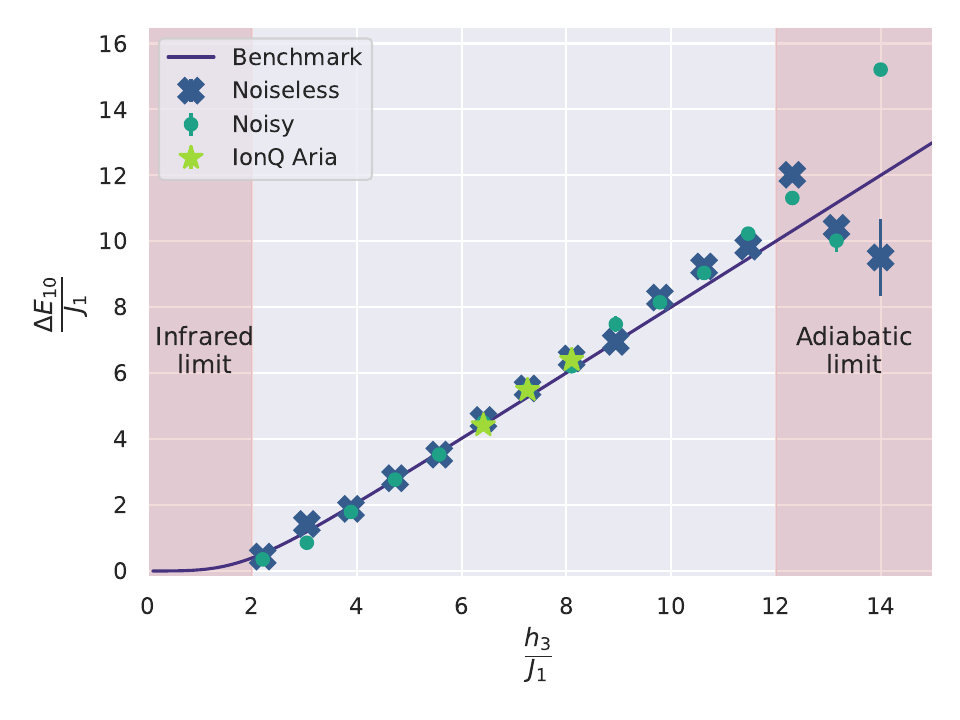}
    \caption{Spectral gap as a function of $h_3/J_1$ for a $L=4$ Ising chain with PBC. The plot includes simulations on a classical computer, showing both noiseless and noisy results, and experiments on IonQ Aria 1. Our approach demonstrates optimal agreement with the benchmark values, highlighting the robustness of our method for execution on NISQ hardware. The adiabatic and infrared limits are regions where the circuit depth used is insufficient. Error bars represent a single standard deviation computed from the fit.}
    \label{fig:spectral gap running}
\end{figure}

The results, shown in Figure~\ref{fig:spectral gap running}, reveal agreement between the noiseless simulation, the noisy simulation, and the IonQ Aria device. This witnesses the robustness of our approach to noise (both simulated and real). It is worth mentioning that the IonQ Aria device comes with integrated error mitigation for a minimum number of 2500 shots, which likely contributed to the consistent results obtained in our study (more details in Appendix~\ref{subsec:ionq}). Moreover, thanks to noise mitigation results on real hardware are actually better than the ones obtained with noisy simulations as they have a bigger $\rho$ (see Appendix~\ref{subsec:ionq}).

\subsection{\ce{H2} and \ce{He2} molecules}
In this section, we present the results of our spectral gap estimation applied to the \ce{H2} and \ce{He2} molecules. 

In general, a molecular Hamiltonian $\mathcal{H}$ is expressed as \cite{moll2016optimizing}
\begin{equation}\label{eq: ce{H2} Hamiltonian}
    \mathcal{H} =  \sum_{pq} h_{pq} c_p^\dagger c_q  + \frac{1}{2}\sum_{pqrs} h_{pqrs} c_p^\dagger c_q^\dagger c_r c_s.
\end{equation}
The coefficients $h_{pq}$ and $h_{pqrs}$,
specific to each molecular system, 
respectively parametrize the one-body and 
two-body electron-electron interactions.
The operators $c_p$ and $c_p^\dagger$ 
represent the fermionic annihilation 
and creation operators acting on an electron 
in the atomic orbital $p$.
The physical Hamiltonian (\ref{eq: ce{H2} Hamiltonian})
can be mapped on N-qubits
via the Jordan-Wigner transformation \cite{shankar2017quantum},
given by the following equation:
\begin{equation}
    H = \Sigma_{i_1, ..., i_L} \left[a_{i_1, ..., i_L}\bigotimes_{j=1}^L\sigma^{i_j}\right],\quad i_j \in \{0,1,2,3\} \ .
\end{equation}
Given the objective to create the  
$\ket{SGS(0,1)}$ state,  
we take the auxiliary Hamiltonian to be 
the diagonal of $H$
\begin{equation}
    H_0 = \mathrm{diag}\left(H\right).
\end{equation}
Since $H_0$ is diagonal,
the eigenstates will
be elements of the computational basis,
making them easy to prepare on
a quantum computer.
In addition, 
one can immediately associate each eigenstate to its eigenvalue.

\begin{figure}
\centering
    \includegraphics[trim={8.5cm 4.5cm 10.3cm 3cm},clip,width=.45\textwidth]{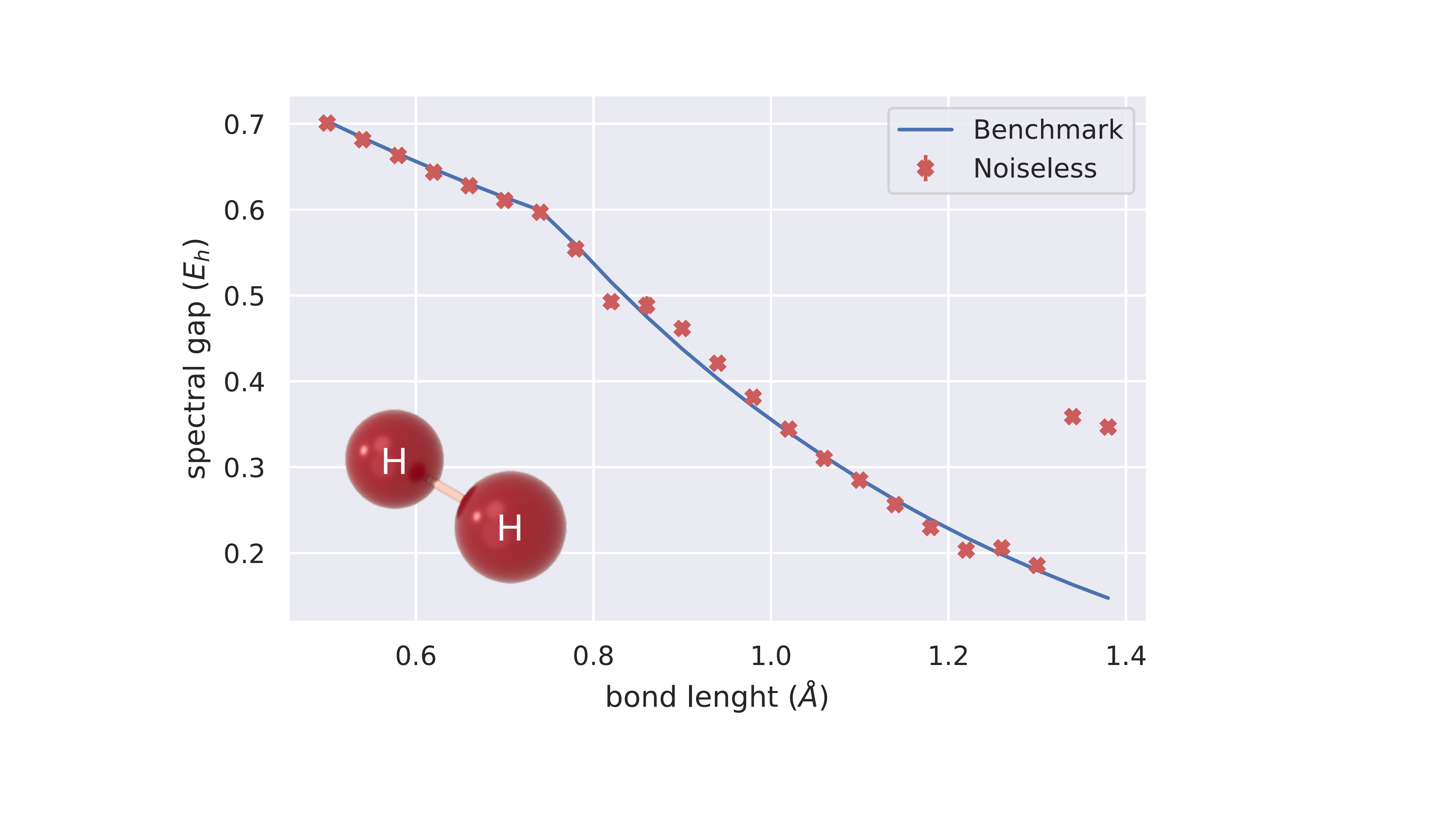}
    \shiftleft{8.4cm}{\raisebox{4.3cm}[0cm][0cm]{(a)}}
    \label{fig: Molecules_H2}
    \includegraphics[trim={10.8cm 4cm 8.cm 5cm},clip,width=.45\textwidth]{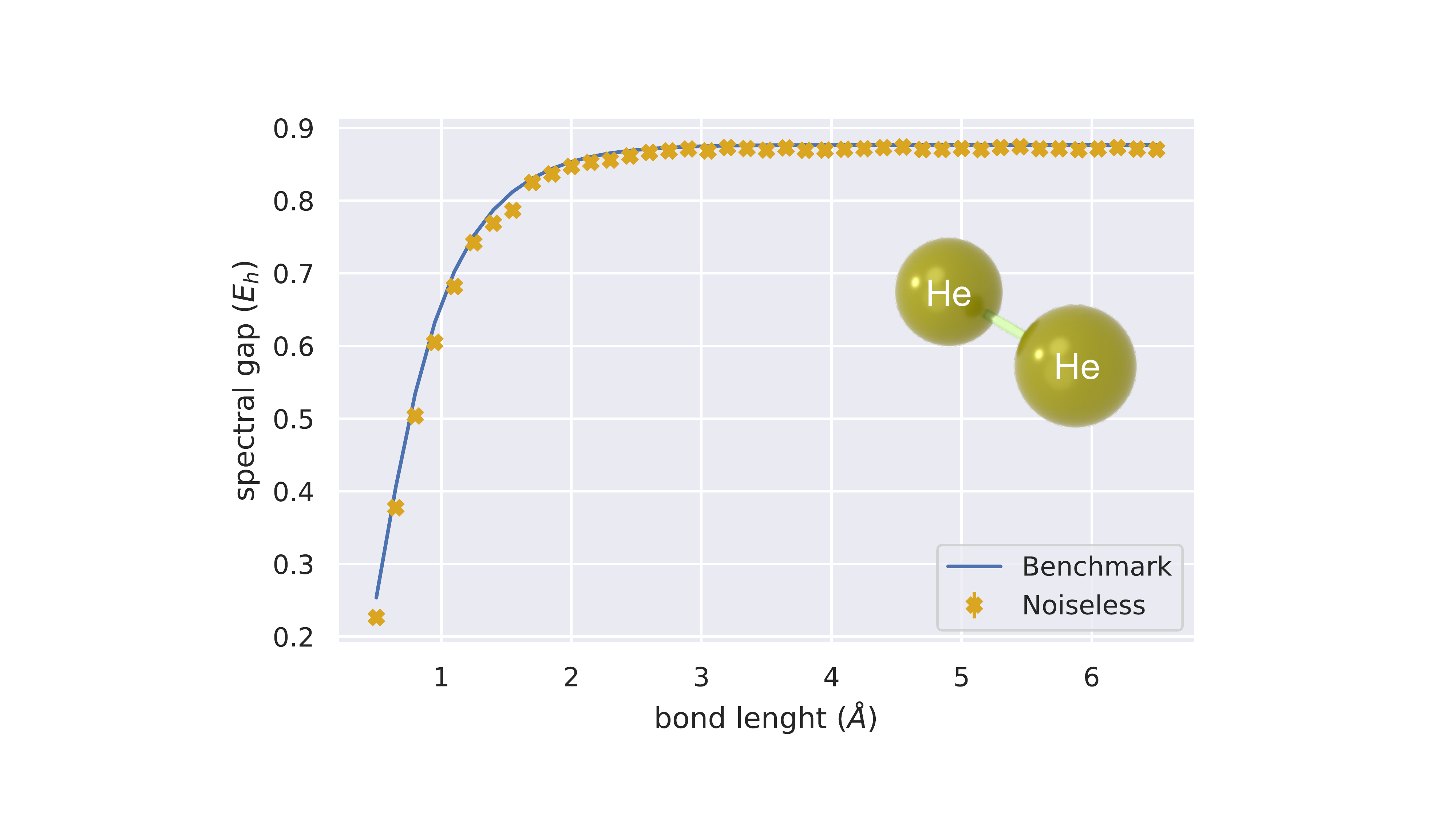}
    \shiftleft{8.4cm}{\raisebox{4.4cm}[0cm][0cm]{(b)}}
    \label{fig: Molecules_He}

\caption{(a) The plot illustrates the spectral gap of the \ce{H2} molecule as a function of the bond length. The noiseless simulation results demonstrate excellent agreement with the benchmark values. (b) The plot illustrates the spectral gap of the \ce{He2} molecule as a function of the bond length. Error bars represent a single standard deviation computed from the fit.\label{fig: Molecules}}

\end{figure}

The state $\ket{SGS_0(0,1)}$ is prepared by
selecting a superposition comprising 
any ground state $\ket{\omega_0}$ and first excited state $\ket{\omega_1}$ of $H_0$. 
Despite the procedure returning accurate
estimates for a wide range of bond lengths,
as seen in Figure~\ref{fig: Molecules},
we believe that symmetry considerations would aid the 
state preparation.

The choice of operators for small 
molecules is fully discussed in Appendix~\ref{sec:observable}. These considerations 
result in an operator $O$ of the form
\begin{equation}
    O = \bigotimes_{j=1}^L \sigma^{i_j}_j \quad i_j \in \{ 0, 1 \}
\end{equation}
which is a tensor product of identities and 
$\sigma^1$ matrices such that:
\begin{equation}
    \abs{\bra{\omega_1} O \ket{\omega_0}} = 1
\end{equation}
where $\ket{\omega_i}$ are the ground and 
excited state of the auxiliary Hamiltonian.

The experiments for the molecules 
involve the calculation of the energy gap 
as a function of the bond length for 
\ce{H2} and \ce{He2}. 
For both experiments 40 Trotter steps
are used, using 5 thermalization steps 
and 35 time evolution steps.
The \ce{H2} Hamiltonian requires 4 qubits,
while the \ce{He2} Hamiltonian requires 8 qubits.

The results from the noiseless simulations
of a \ce{H2} molecule can be seen in  
Figure~\ref{fig: Molecules}a, while those for
\ce{He2} are in Figure~\ref{fig: Molecules}b. The noiseless simulation results demonstrate excellent agreement with the benchmark values, exception made for a couple of points for \ce{H2} possibly due to the choice of the observable, as discussed above.
\\

\section{Discussion}
\label{sec:discussion}
We have developed a novel 
NISQ-friendly procedure to perform energy 
gap estimations for a wide range of Hamiltonians.
Although a procedure that implements time evolution
to obtain information on spectral gaps has been previously attempted ~\cite{GNATENKO2022127843}, our improved method 
is able to specifically estimate any desired energy gap by 
preparing a tailored Spectral Gap 
Superposition state 
exploiting the Adiabatic Thermalization method.
We demonstrated the capabilities 
of our spectral gap estimation procedure 
on the Ising model, both 1D and 2D, as 
well as small molecular systems, \ce{H2}
and \ce{He2}. 
Experiments were conducted using at most 40 
Trotter steps to use circuit depths comparable
with ones available with NISQ devices. 
Our procedure was able to find 
energy gaps with relative errors of the 
order of $10^{-2}$ for noiseless simulations.
The experiments demonstrate 
the procedure holds surprisingly well 
when reducing the number of shots to 100. Furthermore, we were able to 
accurately model the noise of the IonQ 
Aria 1 device.
This led us to launch experiments on the
IonQ Aria 1 device and successfully estimate
the energy gap of a 4-site Ising chain
with a relative error
of the order of $10^{-2}$.
Given the current pace of 
development in quantum computing hardware, 
we believe this procedure could
be used for energy gap estimation  
of bigger molecules and condensed matter systems in the near term. 
In addition, it is worth noting that 
any major breakthrough in either the 
accuracy or the computational cost of 
adiabatic thermalization, which is an active
field of research, will also drastically 
improve the presented energy gap estimation 
procedure. This work opens up the way to effective
quantum computation on NISQ devices for both 
academic and industrial endeavors in the 
realm of Condensed Matter Physics 
and Quantum Chemistry. 

\section*{Acknowledgements}
We would like to thank 
Xanadu Quantum Technologies Inc for
organizing the Pennylane 2024 Coding
Challenges and Quantum Hackathon. 
The presented work was inspired by 
the Hackathon prompt ``Bridging the Gap''. 
Furthermore, we would like to thank 
Amazon Web Services for granting us 
Braket credits to run the experiments 
on real quantum hardware. 

\section{Appendix}
\appendix
\section{Adiabatic Thermalization}
Let $H(t)$ be the time dependent Hamiltonian
\begin{equation}
H(t) =\left( 1 - \frac{t}{\tau}\right)H_0 + \left(\frac{t}{\tau}\right)H
\end{equation}
such that $H(0) = H_0$ and $H(\tau) = H$.
The time evolution operator associated with $H(t)$ is
\begin{equation}
    U_\tau(t) = \mathcal{T}\mathrm{exp}\left[-i\int_0^t ds H(s\tau) \right]
\end{equation}
where $\mathcal{T}$ is the time ordering operator.
In the limit of $\tau \to \infty$ 
the \textit{adiabatic theorem} guarantees that,
under general hypothesis \cite{born1928beweis, jansen2007bounds},
the $n$-th eigenstate $\ket{\omega_n}$ of $H_0$
is evolved by thermalization evolution into
\begin{equation}
    \lim_{\tau \to \infty}U_\tau(1)\ket{\omega_n} = \ket{\Omega_n},
\end{equation}
where $\ket{\Omega_n}$ the $n$-th eigenstate of $H$.
In the main text we use $U_\tau$ to refer to $U_\tau(1)$.

\section{Ising SGS}\label{Ising SGS}
Our aim is to find an initial state that
will guarantee the SGS state as 
defined in Equation~\eqref{eq:SGS} when it undergoes
the thermalization process, 
where we will be switching 
on the magnetic field $h_z$.
To help us in this we introduce the following 
operator:
\begin{equation}
\mathcal{R} = \bigotimes_{j=0}^{L-1} R^{j}_z(\pi)
\end{equation}
where $R_z^j$ is a rotation around the $z$-axis on the $j^{th}$ qubit.
This operator commutes with the Hamiltonian 
for any value of $J$.
This ensures the expectation value
is conserved during the whole AT. 
It is worth noting that the auxiliary 
Hamiltonian ground states, 
$\ket{+}^{\otimes L}$ and $ |-\rangle^{\otimes L}$,
do not diagonalize since
\begin{equation}
\mathcal{R}|\pm\rangle^{\otimes L} = (-i)^L|\mp\rangle^{\otimes L}
\end{equation}

However, it is enough to consider the change of basis
\begin{equation}
|\Phi^\pm\rangle = \frac{1}{\sqrt{2}}\left(|+\rangle^{\otimes L}\pm|-\rangle^{\otimes L} \right)
\end{equation}
to have
\begin{equation}
\mathcal{R}|\Phi^\pm\rangle = \pm(-i)^{L}|\Phi^\pm\rangle.
\end{equation}
We now move our attention to the $\ket{SGS}$ state,
that we want to obtain at the end of the AT,
in the $h_3/J_1 \to \infty $ limit.
In this regime the (non-degenerate) ground state is 
\begin{equation}
|\Omega_0\rangle ^{\otimes L} \equiv \bigotimes_j \frac{1}{\sqrt{2}}\left(|+\rangle +|-\rangle \right)_j
\end{equation}
and respects
\begin{equation}
\mathcal{R} |\Omega_0\rangle ^{\otimes L} = (-i)^L|\Omega_0\rangle ^{\otimes L}.
\end{equation}
The ground state $|0\rangle ^{\otimes L}$ and $\ket{\Phi^+}$ share the same eigenvalue with respect to $\mathcal{R}$, which is preserved during the evolution. It is evident that
\begin{equation}\label{eq: phi+ lim}
\lim_{\tau \to \infty}U_\tau\ket{\Phi^+} =\ket{\Omega_0}
\end{equation}
and, consequently,
\begin{equation}\label{eq: phi- lim}
\lim_{\tau \to \infty}U_\tau\ket{\Phi^-} =\ket{\Omega_1}.
\end{equation}
Moreover, if this is true for $h_3/J_1 \to \infty $
it is true for the cases of finite $h_3/J_1$,
that require a shorter thermalization process.
From Equations~\eqref{eq: phi+ lim} and \eqref{eq: phi- lim}
we finally obtain 

\begin{equation}
    \lim_{\tau \to \infty}U_\tau \ket{+}^{\otimes L} = \ket{SGS(0,1)}.
\end{equation} 

\section{Observable choice}
\label{sec:observable}
For the SGS procedure to work
the objective is to have an 
observable $O$ that maximizes  
$\rho$, as described in Equation~\eqref{eq: fluctuations}. 
This facilitates a more straightforward 
estimation of the period of the 
cosine function, which in turn 
simplifies the fitting process 
for the energy gap. 

\subsubsection*{Ising Model}
For the Ising model our objective 
is to find an operator $O$ that 
maximizes $\rho$ independent 
of the chain/lattice size.
A first set of numerical 
calculations shows that there are 
two operators that achieved this: 
either exactly 1 $\sigma^3$ 
(on any site) and the rest $\sigma^0$,
or exactly 1 $\sigma^1$ 
(on any site) and the rest $\sigma^2$. 
This holds for all Ising chains
with periodic boundary conditions
up to 10 sites. 
Further numerical experiments show
that these same observable 
maximize $\rho$ for Ising lattices 
with up to 9 sites.

Then, we conjecture that the operators 
\begin{equation}
    O_1 = \sigma_{j}^3(\sigma^0)^{\otimes L-1}
\end{equation}
and 
\begin{equation}
    O_2 = \sigma_{j}^1(\sigma^2)^{\otimes L-1}
\end{equation}
maximize $\rho$ as described in equation~\eqref{eq: fluctuations} for an arbitrary Ising 
chain/lattice with $L$-sites, where $O_1$ is a 
$\sigma^3$ on the $j^{th}$
site and the rest $\sigma^0$, 
and $O_2$ is a 
$\sigma^1$ on the $j^{th}$
site and the rest $\sigma^2$.

All our experiments involving the
Ising model use the operator $O_1$ 
to estimate the energy gap. The choice of 
$O_1$ is due to the fact that it is simple
to implement in the circuit.

\subsubsection*{Molecules}
For small molecules the choice of
operator is based on 
the following considerations. 
Since both ground and excited state of 
the auxiliary Hamiltonian are in 
the computational basis,
it is always possible to find a 
tensor product, $P$, of identities and 
$\sigma^1$ Pauli matrices such that
\begin{equation}
    \bra{\omega_1} P \ket{\omega_0} = \bra{\omega_1} \bigotimes_{j=1}^L \sigma^{i_j}_j \ket{\omega_0} = 1, \quad i_j \in \{ 0, 1 \}.
\end{equation}
If the off diagonal terms of $H$ are
small with respect to the diagonal terms,
then choosing $P$ as the operator for
the SGS procedure is unlikely to give $\rho=0$.
In particular
\begin{equation}
    \rho = \abs{\bra{\Omega_1} P \ket{\Omega_0}} = \abs{\bra{\omega_1} U_\tau^\dagger P U_\tau\ket{\omega_0}}
\end{equation}
is unlikely to vanish. 
This is the operator choice for both \ce{H2} 
and \ce{He2} molecules.

\section{Chebyshev nodes}
\label{sec:chebyshev}
The choice of the times used to fit $O(t)$ may be critical,
in the case that they are uniformly distributed, 
because they could accidentally coincide
with the nodes of the cosine function. 
If this was the case, no oscillation of $O(t)$ could be detected. 
Times are chosen as Chebyshev nodes \cite{chebyshev1853theorie}
to avoid this possibility.\\
\section{Intrinsic error estimation}\label{sec: Intrinsic error}
Let
\begin{equation}
\langle O \rangle = \langle \psi |O|   \psi \rangle  
\end{equation}
be the expectation value 
of an operator $O$
on a quantum state $\ket{\psi}$,
and let 
\begin{equation}
\langle\left( O - \langle O \rangle \right)^2 \rangle
\end{equation}
be its variance.
Recalling that $O$ should be unitary and Hermitian 
in order to be measured on a quantum computer 
\begin{equation}
O^2 = O^\dagger O = \mathds{1},
\end{equation}
so that its eigenvalues can only be $\{+1, -1\}$. 
One can then decompose the state 
onto two eigenstates of $O$ as
\begin{equation}
|\psi\rangle  = \alpha|\psi_{+1}\rangle + \beta|\psi_{-1}\rangle
\end{equation}
with
\begin{equation}
\begin{cases}
O|\psi_{\pm1}\rangle = \pm |\psi_{\pm1}\rangle\\
|\alpha|^2+ |\beta|^2 = 1.
\end{cases}
\end{equation}
Then
\begin{equation}
\begin{cases}
\langle O \rangle = |\alpha|^2-|\beta|^2 \\
    \langle\left( O -  \langle O \rangle \right)^2 \rangle = 4|\alpha|^2|\beta|^2
\end{cases}
\end{equation}
\section{More details on IonQ Aria}
\label{subsec:ionq}
\subsubsection{IonQ Native Gates}
\label{sec:native gates}
\begin{figure}[]
    \centering
    \includegraphics[trim={1.cm 1cm 2.cm 1cm},clip,width=.5\textwidth]{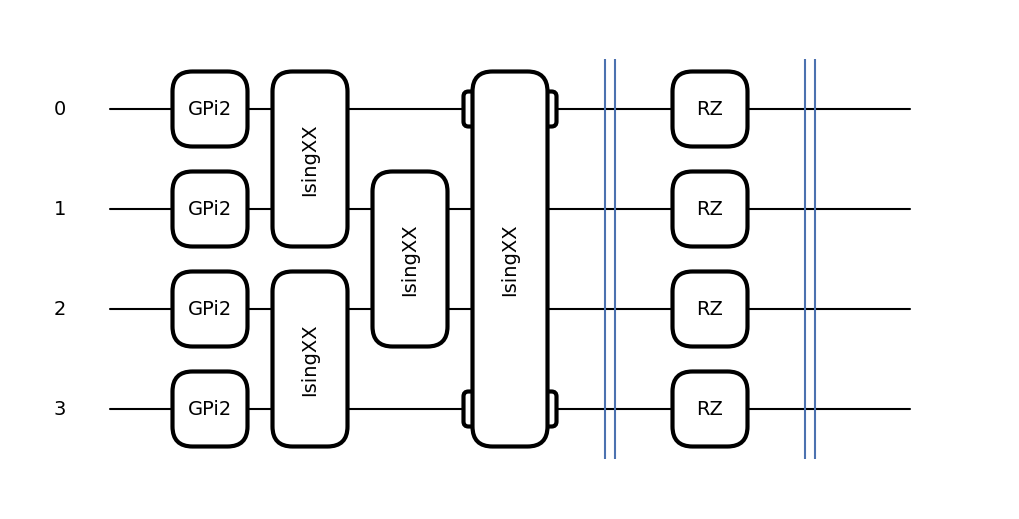}
    \caption{Quantum circuit representation for a single Trotter step illustrating the evolution of the Spectral Gap Superposition (SGS) state. The circuit involves the application of GPI2($\phi$), RZ($\theta$), and IsingXX($\theta$) gates to prepare the SGS state for subsequent spectral gap estimation.}
    \label{fig:trotterstep}
\end{figure}

Since single qubit-gates are typically implemented on ion-trapped hardware with high average gate fidelity, potential issues in result's fidelity (when running a circuit on NISQ hardware) primarily stem from 2-qubit gates. For this reason, we simulate the Ising model with interactions along the $x$-axis and transverse field along the $z$-axis. This allows us to use only the native entangling gates available for IonQ Aria device, i.e. Mølmer-Sørensen gate~\cite{sorensen1999quantum}, and to reduce the number of entangling gates only to the essential ones. More precisely, in a single Trotter step, one only needs two layers of entangling gates (see Figure~\ref{fig:trotterstep}), implying 80 2-qubit gates depth when fixing the number of steps to 40.

The Mølmer-Sørensen gate matrix is 
\begin{widetext}
\begin{equation}
       \text{MS}(\phi_0, \phi_1, \theta) = \begin{pmatrix}
        \cos{\frac{\theta}{2}} & 0 & 0 & -ie^{-i(\phi_0+\phi_1)}\sin{\frac{\theta}{2}}\\
        0 & \cos{\frac{\theta}{2}} & -ie^{-i(\phi_0-\phi_1)}\sin{\frac{\theta}{2}} & 0\\
        0 & -ie^{i(\phi_0-\phi_1)}\sin{\frac{\theta}{2}} & \cos{\frac{\theta}{2}} & 0\\
        -ie^{i(\phi_0+\phi_1)}\sin{\frac{\theta}{2}} & 0 & 0 & \cos{\frac{\theta}{2}}
    \end{pmatrix}.
\label{eq:MS}
\end{equation}
\end{widetext}
This gate matrix can be further simplified to obtain the Ising interaction $\sigma_i^1\sigma_j^1$ by setting $\phi_0$ and $\phi_1$ equal to 0.

\subsubsection{IonQ Aria noise model}
\label{sec:ionq noise}

To assess the resilience of our approach to noise, we incorporated a realistic noise model taking into account the technical features of IonQ's Aria device available online~\cite{ionq_aria}. 
The noise sources considered include:
\begin{itemize}
    \item [-] \textit{Thermal Relaxation}: This source of noise accounts for the effects of the thermal relaxation times $T_1$ and $T_2$. Notably, IonQ Aria has $T_1 = 100s$ and $T_2=1s$, and while we took them into account in our simulations, their effects were found to be negligible and did not significantly impact the results.
    \item [-] \textit{Gate Fidelity}: We simulated errors arising from finite gate fidelity, modeled as depolarizing noise. The relationship between gate fidelity and the depolarizing parameter is derived from what reported in the Appendix of~\cite{blank2020quantum}. The two-qubit gate time is $600\mu s$, while single qubit gates are executed in $135\mu s$.
    \item [-] \textit{Readout Noise} A readout noise of 0.39\% was introduced to emulate imperfections in the measurement process.
\end{itemize}
Upon detailed examination, we have determined that the gate fidelity plays a significant role in the estimation of the spectral gap. This noise source has been implemented by considering a single qubit depolarizing channel, which acts on a density matrix $\Psi$
\begin{equation}
    \mathcal{D}_p (\Psi) = (1-p) \Psi + p\frac{\mathds{1}}{2}\Psi \ .
    \label{eq:depolarizing}
\end{equation}
To take into account gate imperfections and finite qubit relaxation times we define the probability $p$ as
\begin{equation}
    p = 1 + 3 \frac{2\epsilon - 1}{d} \ .
\end{equation}
Here, $d =\exp{-T_g/T_1} + 2\exp{- T_g/T_2}$ and $\epsilon = 1- F$, where $F$ is the average gate fidelity. We define two depolarizing channels one per each type of gate (single/two-qubit gate). 

\begin{figure}
    \centering
    \includegraphics[width=.5\textwidth]{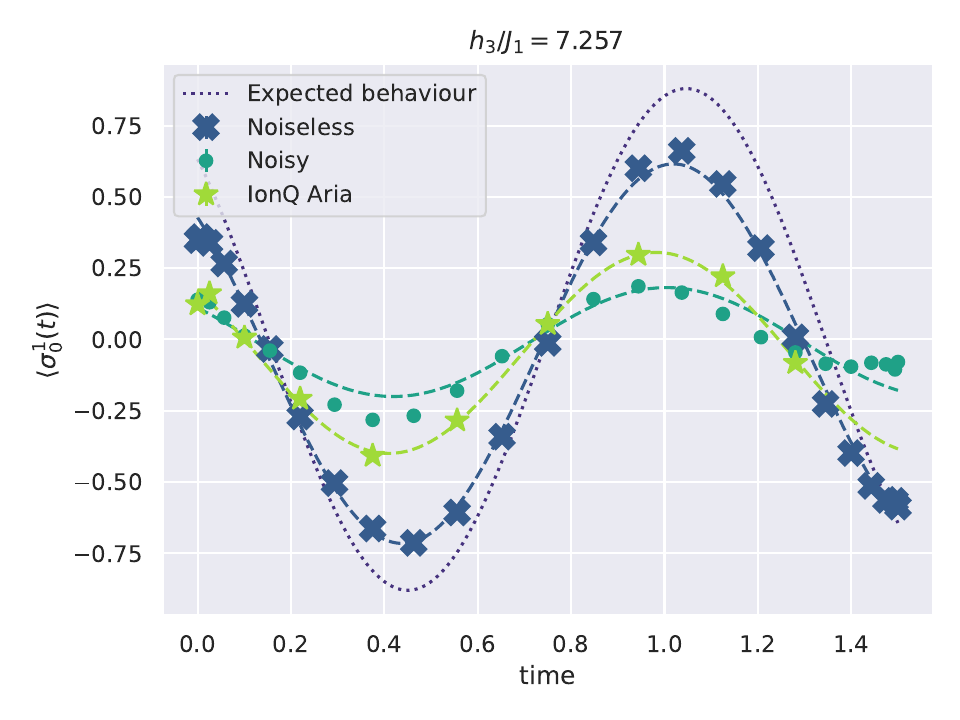}
    \caption{Observable oscillations in time for Ising 1D chain with $L=4$ at $h_3/J_1 = 7.257$, measured (points) and fitted (dashed lines) for noiseless, noisy and real hardware (with error mitigation). For reference, it is also reported the expected oscillatory from exact diagonalization (dotted line). Error bars represent a single standard deviation computed from the fit.}
    \label{fig:waves}
\end{figure}

Our approach demonstrates robustness against gate fidelity errors and readout noise, suggesting its viability for simulation on real quantum hardware.

The noise resilience analysis results are presented in Figure \ref{fig:spectral gap running}. The robustness of our approach can be appreciated also in the time oscillations of the observable reported in Figure~\ref{fig:waves}. There it is highlighted how oscillations are damped by the noise (reduced $\rho$) with respect to the expected behaviour $\rho \cos\left(\Delta E+\theta\right)$. For the expected behaviour we employ $\rho$ given by our preliminary study on the value of the observable and we set the same phase $\theta$ as the noiseless simulations. Interestingly the noise mitigation allows for having wider oscillations on real hardware compared to noisy simulations.

In particular, the IonQ Aria device reduces the impact of systematic errors by employing a specific error mitigation method called \textit{debiasing}~\cite{maksymov2023enhancing}. This technique maps a circuit into multiple variants, employing different qubit permutations or gate decompositions.

\begin{figure}[t]
    \centering
    \includegraphics[width=.45\textwidth]{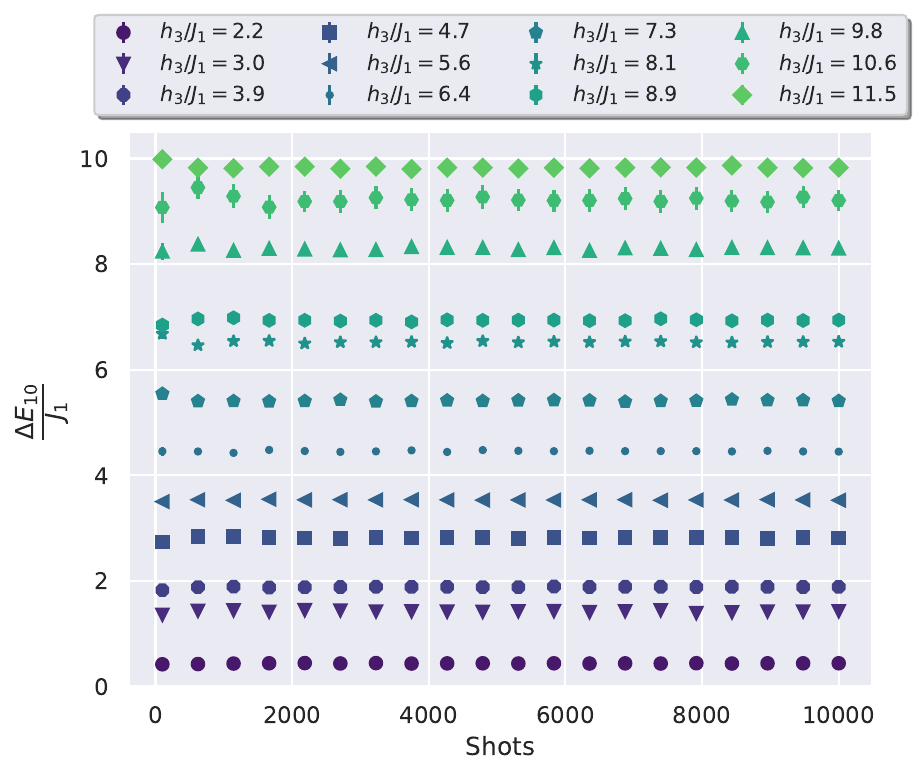}
    \caption{Robustness testing over increasing number of shots. Scatter points correspond to measured value of the spectral gap in $J_1$ units. Error bars represent a single standard deviation computed from the fit.}
    \label{fig:shots noise}
\end{figure}

The effectiveness of debiasing error mitigation is likely improved by our approach's robustness to a reduced number of shots. In fact, as reported in Figure~\ref{fig:shots noise}, we assess the robustness of our approach over a reduced number of shots by reducing this value up to 100 shots. As witnessed by Figure~\ref{fig:shots noise}, the spectral gap estimate stays approximately constant over the different number of shots.

\begin{figure}[t!]
    \centering
    \includegraphics[trim={.3cm 0.cm 0.cm 0cm},clip,width=0.45\textwidth]{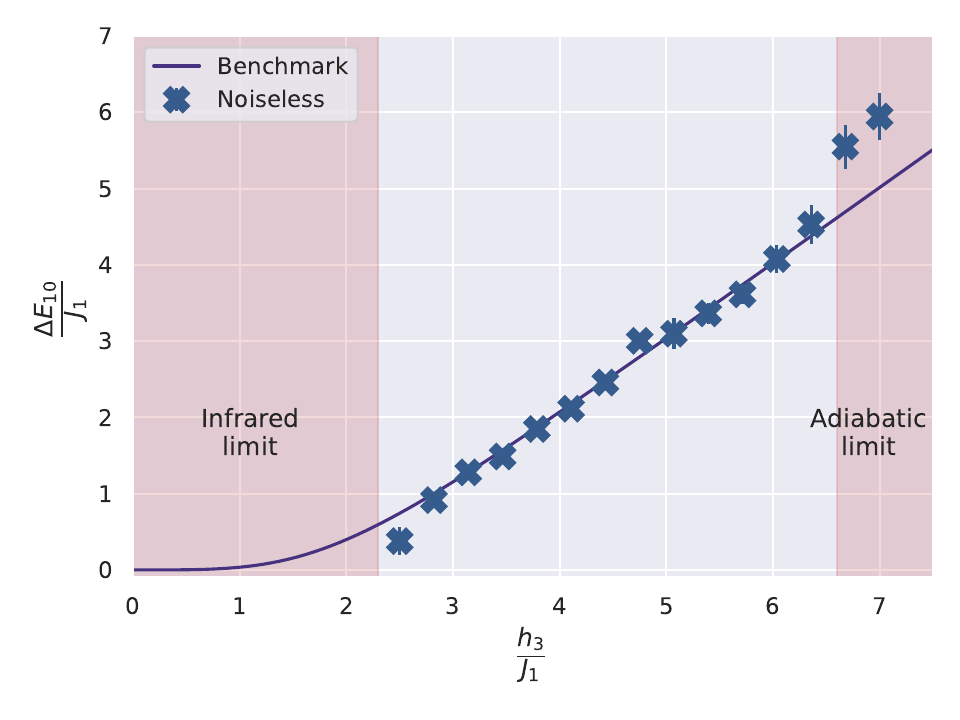}
    \caption{Spectral gap as a function of $h_3/J_1$ for a $L=4$ Ising 2D lattice with periodic boundary conditions (PBC). Our approach demonstrates optimal agreement with the benchmark values, in the noiseless case. Error bars represent a single standard deviation computed from the fit.}
    \label{fig:Ising 2D}
\end{figure}

\section{Ising Lattice experiments}
\label{subsec:Ising2d}
Here we present further experiments related to 
an Ising lattice with 4 sites 
and periodic boundary conditions (see Figure 
\ref{fig:Ising 2D}). 
All previous conditions for the Ising chain
experiment remain in this study. 
In particular, the 40 Trotter steps are divided in 15 thermalization
steps and 25 time evolution steps. In the 2D simulations, the number of entangling gates required per Trotter step is doubled with respect to the 1D model. Due to the prohibitive depth implied by the extra interactions, simulations with real hardware noise model lead to inconclusive results. It would be interesting to see if IonQ Aria noise mitigation is able to lead to measurable results.

\bibliography{bibliography}

\end{document}